\documentclass[english]{article}
\usepackage[T1]{fontenc}
\usepackage[latin9]{inputenc}
\usepackage{array}
\usepackage{amstext}

\makeatletter

\providecommand*{\perispomeni}{\char126}
\AtBeginDocument{\DeclareRobustCommand{\greektext}{%
  \fontencoding{LGR}\selectfont\def\encodingdefault{LGR}%
  \renewcommand{\~}{\perispomeni}%
}}
\DeclareRobustCommand{\textgreek}[1]{\leavevmode{\greektext #1}}
\DeclareFontEncoding{LGR}{}{}

\newcommand{\lyxmathsym}[1]{\ifmmode\begingroup\def\b@ld{bold}
  \text{\ifx\math@version\b@ld\bfseries\fi#1}\endgroup\else#1\fi}

\providecommand{\tabularnewline}{\\}

\newcommand{\lyxaddress}[1]{
\par {\raggedright #1
\vspace{1.4em}
\noindent\par}
}

\usepackage{color}
\usepackage{soul}

\makeatother

\usepackage{babel}

\begin{document}

\title{De retibus socialibus et legibus momenti%
\thanks{On social networks and the laws of influence%
}}

\author{Daniel Gayo-Avello%
\thanks{Correspondence to: Daniel Gayo-Avello, Department of Computer Science
(University of Oviedo) -- Edificio de Ciencias, C/Calvo Sotelo s/n
33007 Oviedo (SPAIN), \texttt{dani@uniovi.es}%
}, David J. Brenes, Diego Fernández-Fernández,\\
María E. Fernández-Menéndez, Rodrigo García-Suárez}

\maketitle

\lyxaddress{\noindent \begin{center}
University of Oviedo (SPAIN) \& Simplelogica, SL (SPAIN)
\par\end{center}}
\begin{abstract}
\emph{Online Social Networks }(\emph{OSNs}) are a cutting edge topic.
Almost everybody --users, marketers, brands, companies, and researchers--
is approaching \emph{OSNs} to better understand them and take advantage
of their benefits. Maybe one of the key concepts underlying \emph{OSNs}
is that of influence which is highly related, although not entirely
identical, to those of popularity and centrality. Influence is, according
to Merriam-Webster, \emph{{}``the capacity of causing an effect in
indirect or intangible ways''}. Hence, in the context of \emph{OSNs},
it has been proposed to analyze the clicks received by promoted \emph{URLs}
in order to check for any positive correlation between the number
of visits and different {}``influence'' scores. Such an evaluation
methodology is used in this paper to compare a number of those techniques
with a new method firstly described here. That new method is a simple
and rather elegant solution which tackles with \emph{influence} in
\emph{OSNs} by applying a physical metaphor.
\end{abstract}

\section*{Introduction}

This paper describes an eminently empirical study for which a number
of experiments were conducted. \emph{Twitter} was chosen for that
purpose because it is relatively easy to obtain data from it in comparison
to other services such as \emph{Facebook}. For those experiments the
\emph{Twitter} dataset from \cite{key-6} was used. That study completely
describes the data assets but, still, a brief description appears
at the end of this introductory section.

\subsection*{Basic concepts}

\emph{Twitter }is a \emph{microblogging} service which allows users
to publish text messages of up to 140 characters (\emph{tweets}) which
are shown to other users subscribed to the author feed (\emph{followers}).
Unlike other \emph{OSNs}, relationships in \emph{Twitter} are asymmetrical
and, thus, it must be distinguished between people reading a given
author messages (the aforementioned \emph{followers}), and those persons
that author reads (\emph{friends }or \emph{followees}).

\subsection*{The user graph and eigenvector centrality}

Therefore, \emph{Twitter }can be represented as a directed graph and,
hence, it is amenable to analysis by means of eigenvector centrality
algorithms. The aim of such algorithms is to compute the centrality
of a node within a network (i.e. a graph) starting from rather simple
assumptions: \emph{(1)} the centrality of a node depends on the respective
centrality values of the nodes linking to it; \emph{(2)} the more
nodes linking to a given one, or the more central the few nodes linking
to it, the more central that node will be; \emph{(3)} centrality values
for all of the nodes within the network are iteratively computed until
the algorithm converges. 

Nonetheless to say, a number of algorithms exist to compute one or
another {}``flavor'' of these centrality scores. The {}``power
iteration'' method to compute the eigenvalues and eigenvectors of
a matrix $M$ is one of such methods. \emph{PageRank} \cite{key-2},
\emph{HITS} \cite{key-3}, \emph{TwitterRank} \cite{key-4}, or \emph{TunkRank}
\cite{key-5} are other approaches to compute slightly different scores
better adapted to the graph properties of the WWW or the \emph{Twitter}
user graph.

The interested reader is recommended to consult \cite{key-6} for
a deep study and comparison of such algorithms regarding their application
to the \emph{Twitter} user graph. Suffice to say here that centrality
algorithms are sensitive to a common form of abuse in \emph{Twitter}
--the \emph{follow-to-be-followed} pattern-- and, thus, robust methods
to compute centrality are needed, in addition to verifying whether
centrality is actually related or not to the elusive \emph{influence}.

\subsection*{A naïve approach to popularity and influence}

The number of \emph{followers} has been largely considered as equivalent
to \emph{popularity}. After all, it seems rather obvious that the
more \emph{followers} a user has got the more popular s/he is and,
in fact, celebrities such as Lady Gaga or Britney Spears have got
millions of \emph{followers}. 

Given this simple approach to \emph{popularity}, many \emph{Twitter
}users have exploited a simple rule of etiquette to get massive audiences.
In \emph{Twitter}, it is considered good manners to \emph{follow back}
a new \emph{follower} and, hence, some abusive users (such as spammers
and aggressive marketers) tend to follow thousands of people to get
\emph{followers} in return. 

Because of this behavior, the \emph{followers/followees} ratio has
also been used as a proxy measure for a user's \emph{influence}: those
users with a ratio greater than 1 are {}``influential'' while those
with a ratio lesser than 1 are {}``uninfluential''; besides, the
larger the ratio the more {}``influential'' the user.

\subsection*{Users' tweeting behavior}

In addition to \emph{following} behaviors, \emph{Twitter} users also
get involved in \emph{tweeting} behaviors: thus, a \emph{tweet }can
be original content produced by the author or it can be non-original
content; that is, the user is repeating a \emph{tweet} by another
user (\emph{retweeting}, in \emph{Twitter} parlance). 

Because \emph{retweeting} is a form of citation, some syntax to provide
attribution is needed. To do that, the name of the mentioned user
is prepended with an \emph{at sign}. 

Let's suppose, for instance, that Alice had \emph{tweeted} the message
\texttt{{}``Hello world!''}. If Bob wanted to \emph{retweet} it
he just should have to post%
\footnote{It must be noted that users in \emph{Twitter} do not necessarily use
their real name as user name. For instance, Ashton Kutcher is \texttt{aplusk}
and CNN Breaking News is \texttt{cnnbrk}.%
}: \texttt{{}``RT @alice: Hello world!''}, where \emph{RT} stands
for \emph{retweet}. 

Of course, this \emph{mention} syntax is not limited to \emph{retweets}
and, indeed, it can be used to address other users and get involved
into conversations rather similar to those within \emph{IRC} (\emph{Internet
Relay Chat}). 

So, in short, users can \emph{tweet}, \emph{retweet}, \emph{mention},
or combine all of them --e.g. \emph{retweeting} some content while
addressing it to a third party: \texttt{{}``RT @alice: Hello world! (cc
@carol)''}. For a deeper understanding of the \emph{tweeting} and
\emph{retweeting} behavior of users we highly recommend the work by
boyd \emph{et al.} \cite{key-11}.

Last, but not least, every \emph{tweet} is timestamped and, therefore,
it is possible to compute for every user his \emph{tweeting} frequency,
bursts of activity, idle periods, etc.

\subsection*{Other approaches --different from eigenvector centrality-- to compute
influence}

\subsubsection*{Feature rich approaches}

Thus, \emph{Twitter} provides plentiful of user features: \emph{i)}
number of \emph{followers} and \emph{followees}; \emph{ii)} number
of \emph{tweets}, \emph{retweets}, and \emph{mentions} --both produced
and received; \emph{iii)} total number of \emph{tweets}, and average
number of \emph{tweets} per hour, day, or week, etc. 

All of these features are being used in almost any conceivable combination
to produce formulae to score \emph{Twitter} users. Companies such
as \emph{Klout}, \emph{PeerIndex}, \emph{tweetreach}, or \emph{Twitalyzer}%
\footnote{\texttt{http://klout.com}, \texttt{http://www.peerindex.net}, \texttt{http://tweetreach.com},
and \texttt{http://twitalyzer.com}.%
} use them to compute \emph{ad hoc} scores which, arguably, provide
a glimpse into the \emph{influence} or \emph{authority} of a given
\emph{Twitter} user.

Needless to say, actual details for such scores are undisclosed. Nevertheless,
the interested reader can consult the recent work by Pal and Counts
\cite{key-7} describing a method to \emph{(1)} cluster \emph{Twitter}
users according to several features --including those aforementioned,
and \emph{(2)} rank users within the found clusters to find topical
experts.

\subsubsection*{Influence maximization and diffusion cascades}

A different angle of approach has been inspired by the highly influential
work by Domingos and Richardson \cite{key-8}, and Kempe \emph{et
al.} \cite{key-9}. Simply put, these researchers have studied the
way in which \emph{influence} (e.g. related to purchase intention)
\emph{virally} spreads through users within a network, so a minimum
set of \emph{influential} users can be found (i.e. those that should
be addressed by a marketer in order to maximize sales with minimum
effort). 

The so-called \emph{diffusion cascade models} --which are highly related
to this area of \emph{influence maximization}-- have been rather successfully
applied to \emph{Twitter }(e.g. \cite{key-1,key-10,key-12}).

It should be noted, however, that finding the optimum for influence
maximization is NP-hard and, thus, efficient approximate algorithms
are used. In this sense, Java \emph{et al.} \cite{key-14} showed
that \emph{PageRank} can be a feasible and inexpensive solution. Therefore,
in spite of being a different approach, eigenvector centrality seems
to be a good approximation to influence maximization in \emph{OSNs}
for all practical purposes.

\subsubsection*{The Influence-Passivity method}

Finally, two recent works by Huberman \emph{et al.} \cite{key-15},
and Romero \emph{et al.} \cite{key-17} must be cited. The former
revealed important differences between the \emph{Twitter }{}``declared''
user graph and the actual interaction graph which is, in some sense,
{}``hidden''. The {}``declared'' user graph is built from the
\emph{follower-followee} relationships between users. The interaction
graph, instead, does not take into consideration all of these relationships
but only those which also involve actual interactions (i.e. users
\emph{mentioning} each other). Such a graph is {}``hidden'' because
interactions are not part of the user graph but have, instead, to
be inferred from the \emph{tweet} streamline.

The implications of this are clear: first, the number of \emph{followers}
and \emph{followees} are misleading if there are no actual interaction
between users; second, centrality measures obtained from the {}``declared''
user graph could be very different from those obtained from the {}``hidden''
interaction graph. 

The second work is highly related to the first one; in it Romero \emph{et
al.} described the so-called \emph{Influence-Passivity }algorithm.
In certain sense this new algorithm is closely related to others such
as \emph{PageRank}, \emph{HITS}, or \emph{TunkRank}. However, unlike
them, the edges (their weights, indeed) and partial scores are inferred
from user interactions, in concrete, \emph{retweets}. 

The assumptions underlying their approach are very appealing: \emph{(1)}
The \emph{influence }of a user depends on the \emph{passivity} of
his \emph{followers} and, conversely, the \emph{passivity} depends
on the influence of his \emph{followees}. \emph{(2) }For each pair
of users, an \emph{acceptance} and a \emph{rejection} rate are computed
for the \emph{follower} user. The former is the amount of \emph{influence
}the \emph{follower} accepts from his \emph{followee} (i.e. the number
of received messages s/he \emph{retweets}) while the later is the
amount of \emph{influence} the \emph{follower} \emph{rejects}. \emph{(3)}
This way, the passivity of a user is proportional to both his \emph{rejection
rate} and the influence of his \emph{followees}%
\footnote{That is, a user rejecting \emph{tweets} from more influential users
is more \emph{passive} than a user rejecting the same amount but from
less influential \emph{followees}.%
} while the \emph{influence} of a user is proportional to both the
\emph{acceptance rate} and the \emph{passivity} of his \emph{followers}.
Finally, \emph{(4) influence} and \emph{passivity} scores are computed
with an iterative algorithm that converges in relatively short time.

\subsection*{Dataset acquisition and description}

The dataset used in this study consists of a collection of 27.9 million
\emph{tweets} and a user graph comprising 1.8 million users. Both
were obtained using a number of methods of the \emph{Twitter API (Application
Programming Interface)}. The \emph{tweets} were collected from January
26 to August 31, 2009. Due to some network blackouts 4 days are missing
and, thus, the dataset has got, on average, 130,000 \emph{tweets}
per day. On 2009 \emph{Twitter} received 2.5 million \emph{tweets}
per day \cite{key-1}, hence, the data corresponds to about 5.6\%
of the total amount of \emph{tweets} published during the crawling
period.

\emph{Tweets} are associated with metadata such as the publishing
author and, thus, a list of 4.98 million users was obtained from the
previous dataset and used to crawl the user graph. At the moment of
that second crawling many accounts had been \emph{suspended} or changed
their status from public to private. Additionally, users without links
to other users in the list were considered isolated and removed, and
there were also minor network blackouts. For these reasons the graph
contains less users than those publishing \emph{tweets}. Anyhow, it
was checked that the crawling was uniform and, in fact, the graph
corresponds to 4\% of the \emph{Twitter}'s worldwide user base of
44.5 million users as of mid-2009 \cite{key-2}.

\section*{A proposal for Twitter dynamics}

\subsection*{Rationale}

It is clear that to apply any of the above-mentioned methods to compute
\emph{influence} in \emph{Twitter} the user graph is needed. That
graph alone is enough for eigenvector centrality methods but for the
rest of approaches the published \emph{tweets} are also required.
Such data is needed to find out the \emph{retweets}, \emph{mentions},
diffusion cascades, and {}``hidden'' relations between users. 

Thus, researchers and practitioners working with \emph{Twitter} usually
deal with both data assets. It should be noted, however, that these
two kinds of data (\emph{tweets} and user graph) are not only distinct
in nature but they are also crawled in very different ways. 

\emph{Tweets} can be relatively easy obtained as a data stream and
most of the computation on them can be performed in near real-time.
The user graph, however, is a \emph{snapshot} taken at a given time
or, at most, a series of periodic \emph{snapshots}. Nevertheless,
\emph{Twitter }in particular, and \emph{OSNs} in general are highly
dynamic systems, with users joining and quitting the network, and
linking and unlinking among them continuously. Thus, static \emph{snapshots}
are a pale approximation to the actual evolving network. 

Of course, it can be argued that a reasonable approximation is better
than no approximation at all; however, in the light of recent findings
such as those by Huberman \emph{et al.} \cite{key-15}, we should
wonder: \emph{Is the user graph really needed to get a picture of
Twitter?} Even more concretely, \emph{is there any way of inferring
influence by just relying on the most basic actions of Twitter users?}

The method described in the following subsection demonstrates that
the user graph can be greatly disregarded, and \emph{mentions} are
enough to provide not only an accurate picture of \emph{Twitter} but
a dynamic one. Given that \emph{mentions} are citations this should
be hardly unsurprising; however, our approach is not based on bibliometrics
but on physics, concretely on dynamic friction and uniformly accelerated
linear motion.

\subsection*{A physical metaphor for influence in Twitter}

The approach here described is an answer to the two aforementioned
research questions and it evolved after a number of iterations.

Firstly, the role of the user graph to determine user \emph{influence}
was debated: the user graph could be \emph{(a)} an essential data
asset as in the cases of \emph{PageRank} and \emph{TunkRank}; \emph{(b)}
a starting point to find out the actual interaction graph as in the
work by Huberman \emph{et al}.; or \emph{(c)} a dispensable asset.
It is rather obvious that in the later case user interactions would
be the only data to work on --no matter whether or not there were
any \emph{follower-followee} relationship between users.

Still, it was possible to build an implicit user graph from such users
interactions in such a way that any graph-based method could be applicable.
However, not building such an implicit graph was not only a novel
approach but, besides, it would make real-time computations easier.
Therefore, it was decided to study such an approach.

By totally disregarding both explicit and implicit user graphs it
was clear that user influence would mostly rely on the \emph{mentions}
received by the users. It was also clear that a mere accounting of
the total number of \emph{mentions} received could be as misleading
as the \emph{follower} count. That is, it would not take account of
the dynamic nature of \emph{mentions} as they are related to events
in which the mentioned user is involved. Hence, the time factor should
be accounted for and, that way, it was perceived that equally or more
important than the number of \emph{mentions} would be the rate with
which \emph{mentions} increase. It was in that moment that the similarities
to the dynamics equations were noted, and it was decided to study
the feasibility of adapting them to compute users' \emph{influence}
in \emph{OSNs}, concretely in \emph{Twitter}. 

Hence, in short, to devise this new approach, concepts from dynamics
such as \emph{force}, \emph{mass}, \emph{acceleration} and \emph{velocity}
have been translated to an \emph{OSN} scenario%
\footnote{We'd like to say that this is the first time that such a physical
approach is suggested for \emph{OSNs}; however, on October 20, 2010
the so-called \emph{{}``velocity and acceleration''} model was reported.
It must be said that, unlike our method, such a model is not an adaptation
of physical laws but, instead, \emph{velocity} and \emph{acceleration}
are used to denote the first and second derivatives of the time series
corresponding to the tweet volume for a given topic \cite{key-3}.
Needless to say, both derivatives provide interesting information
about the shape of the curve and, thus, the behavior of the topic
but they are not a proper physical model and, hence, both their method
and ours are unrelated.%
}. Thus, a user's \emph{influence} is modeled as \emph{velocity} and,
thus, \emph{acceleration} can be used to detect trending users in
real time. 

Let's start with Newton's second law: 

\[
F=m\text{·}a\]

How does this translate to \emph{Twitter}? First, the \emph{mass }of
a user is the number of \emph{followers}. Second, the \emph{force}
applied to put a user {}``in motion'' is the number of \emph{mentions}
received (\emph{retweets} alone could also be used). This way, a user
with a high number of \emph{followers} needs more \emph{mentions}
to start {}``moving'' while a {}``lighter'' user (one with a lower
number of \emph{followers}) requires fewer \emph{mentions}. 

It should be noted, however, that this equation assumes instantaneous
\emph{forces} and \emph{accelerations}, and continuous time. For implementation
purposes it is much more simple, however, to operate in discrete time
intervals. Therefore, all of the experiments described in this paper
were performed using one-hour sampling intervals. This way, the \emph{force}
applied on a user is, indeed, the number of \emph{mentions} addressing
that user in a given hour%
\footnote{That sampling interval was fixed after some proof of concept experiments.
When using shorter intervals (e.g. one minute) most of the users did
not receive any mention, and when receiving any the {}``applied force''
was virtually negligible. Larger intervals (e.g. one day) solved that
problem but they were too coarse-grained for events evolving along
hours.%
}.

In addition to this, under real circumstances there are more forces
at stake: mainly, the force of kinetic friction \textbf{$F_{f}$}.
Thus, \emph{mentions} are actually the applied force, \textbf{$F_{a}$},
while \textbf{$F$} is the resultant force of \textbf{$F_{a}$} and
$F_{f}$. The equation for the force of kinetic friction is the following:

\[
F_{\text{f}}=\mu\text{·}N\]

Where $N$ is the normal force and $\lyxmathsym{\textgreek{m}}$ the
coefficient of friction. That way the \emph{acceleration} would be:

\[
a=\frac{F_{\text{a}}-\mu\text{·}N}{m}=\frac{F_{\text{a}}-\mu\text{·}m\text{·}g\text{·}cos(\Theta)}{m}=\frac{F_{\text{a}}}{m}-\mu\text{·}g\text{·}cos(\Theta)\]

Because the equation is to be translated to a non-physical scenario
it can be simplified by supposing that not only $\lyxmathsym{\textgreek{m}}$,
and $g$, but also \emph{$cos(\Theta)$} are constant for every run
of the method; thus, \emph{acceleration }in \emph{Twitter} would actually
be:

\[
a=\frac{F_{\text{a}}}{m}-\zeta\]

Where $\lyxmathsym{\textgreek{z}}$ is a damping constant which is
responsible for the decay of users' \emph{acceleration} and \emph{velocity}
when they do not receive any \emph{mention}. Needless to say, the
value for that constant must be empirically determined and should
have the same dimensions as the quotient $\frac{F_{\text{a}}}{m}$,
that is, \emph{mentions} per hour per \emph{follower}. Hence, $\lyxmathsym{\textgreek{z}}$
value would be the average number of \emph{mentions} per hour per
user, divided by the number of \emph{followers} an average user has
got in \emph{Twitter}.

Finally, the \emph{velocity} of a \emph{Twitter} user would be computed
according to the following equation:

\[
v_{\text{t}}=v_{\text{t-1}}+\frac{F_{\text{a}}}{m}-\zeta\]

It must be taken into account that \emph{(1)} time is discrete, using
one-hour sampling; \emph{(2)} \emph{$m$} is the number%
\footnote{In fact, smoother results can be obtained by applying natural logarithms
to the number of \emph{followers}.%
} of \emph{followers} of the user; \emph{(3)} $F_{a}$ is the number
of \emph{mentions} addressing the user in the last hour; \emph{(4)}
$\lyxmathsym{\textgreek{z}}$ is a constant positive number; and \emph{(5)}
negative velocities are not allowed and, hence, they should be replaced
by zero.

From this equation it is easy to see that a frictionless scenario
($\lyxmathsym{\textgreek{z}}=0$) is a special case where \emph{velocity}
is the accumulated number of \emph{mentions} received by a user divided
by his number of \emph{followers}. Besides, if all of the users had
the same number of \emph{followers} then \emph{velocity} would be
equivalent to citation count.

In addition to that, by knowing both \emph{velocity} and \emph{acceleration}
for each user at every hour it would be possible not only to know
users \emph{influence} but, much more importantly, to find \emph{trending}
users --i.e. those with higher \emph{accelerations}-- in real-time.
Anecdotal evidence on this point is provided in a later section.

\section*{Experimental evaluation}

\subsection*{Influence\ensuremath{\approx}Attention\ensuremath{\approx}Clicks}

So far, another model to compute a score which may or many not relate
to \emph{influence} has been proposed. Thus, a way to correlate \emph{velocity}
with \emph{influence} was also needed. 

As it has been said, \emph{influence} should exert measurable effects
and, in this sense, the evaluation approach by Romero \emph{et al.}
\cite{key-17} is pretty clever: they argued that, in the context
of \emph{Twitter}, \emph{influence} should correlate with attention
and, therefore, \emph{URLs} posted by influential \emph{Twitter} users
should receive more visits than those \emph{URLs} published by less
influential ones.

Needless to say, the number of visits a given \emph{URL} receives
is just known to each website administrator. However, because of the
length limit of the \emph{tweets} (140 characters at most) virtually
every published \emph{URL} is \emph{shortened} by means of one of
several services%
\footnote{Using a shortening service a \emph{URL} such as \texttt{http://en.wikipedia.org/wiki/}~\\
\texttt{URL\_shortening} translates into \texttt{http://bit.ly/ebfVuu}.%
}. 

One of them, \emph{bit.ly}, provides an \emph{API} which allows anyone
to check the number of \emph{clicks} a given \emph{short URL} has
received. Hence, using that \emph{API}, it was possible to associate
to \emph{bit.ly} \emph{URLs} appearing in \emph{tweets }from the aforementioned
dataset the corresponding number of visits those \emph{URLs} had received%
\footnote{Not every \emph{bit.ly URL} appearing in the dataset was used, only
those which were published by at least 3 users for whom graph data
(i.e. their \emph{followers} and \emph{followees}) was available and
it was possible to compute the \emph{I-P} score by means of the \emph{Influence-Passivity}
method.%
}. Then, it was quite straightforward to check for any correlation
between the \emph{influence }of the users publishing the \emph{URLs}
and the visits for those \emph{URLs}.

It must be noted, however, that some changes were made to the data
collection methodology by Romero \emph{et al}. They worked with \emph{URLs}
without taking into account for how long such \emph{URLs} appeared
in the \emph{Twitter} stream. This is quite pertinent because some
\emph{URLs} can consistently appear for weeks or months, achieving
a high number of visits with little or no relation at all with the
\emph{influence} of the users posting them%
\footnote{For instance, \emph{URLs} such as \texttt{http://bit.ly/SXp2X} or
\texttt{http://bit.ly/2MbrXo} appeared virtually every week in our
dataset; as it can be easily checked they are horoscopes. It is obvious
that these are not the only websites that can recurrently appear in
the \emph{Twitter} stream (think for instance of news, auctions, or
music sites).%
}. 

Therefore, in addition to preparing a \emph{URL} dataset in the fashion
of Romero \emph{et al.}, a second one comprised of \emph{URLs} appearing
during one single week was prepared. An additional advantage of this
second dataset is that it made possible to correlate \emph{URL} visits
with \emph{velocity} values computed each week instead of comparing
visits with one single final score for each user. 

Finally, outliers were eliminated from both datasets using the common
interquartile range method ($k=1.5$). To that end, URL visits were
considered and those URLs with exceedingly high numbers of visits
were removed. In the second dataset, the outliers were computed for
each different week and not for the whole dataset. 

Hence, the first dataset was finally comprised of 22,920 \emph{URLs}
while the second one contained 10,120 \emph{URLs} distributed in 29
weeks --from January 26 to August 16, 2009-- with an average of 349
\emph{URLs} and a standard deviation of 139.4.

\subsection*{Influence metrics evaluated}

Romero \emph{et al. }compared the predictive power of their \emph{Influence-Passivity
(I-P)} score with both \emph{PageRank} and the number of \emph{followers}.
For this study not only those metrics were compared but also the recently
proposed \emph{TunkRank}, and the new method described in this paper
--i.e. \emph{velocity}.

Therefore, the number of \emph{followers}, \emph{I-P}, \emph{PageRank},
and \emph{TunkRank} were determined for those users appearing in the
\emph{Twitter }dataset described in\textbf{ }\cite{key-6}. In addition
to that, \emph{velocities} were computed and those reached at the
end of each week were stored. 

Needless to say, it was not possible to compute all of the scores
for every user in the dataset: \emph{(1)} \emph{PageRank} and \emph{TunkRank}
require graph data for the users. \emph{(2)} \emph{Velocity} requires
the users to be mentioned in the \emph{tweets}. And \emph{(3)}, \emph{I-P}
does not only require graph data but also that connected users are
involved in \emph{retweeting} behavior. Thus, only those users qualifying
for all of the methods were considered for the experiments%
\footnote{These meant about 12,000 users; the strict requirements of the \emph{Influence-Passivity}
method --i.e. to just consider connected users who also \emph{retweet}
each other-- drive to a very sparse graph.%
}.

That way it was possible to associate every \emph{URL} with both a
number of visits, and a list of users who had {}``promoted'' that
\emph{URL} in \emph{Twitter}. Those users, in turn, had known {}``influence''
scores. Therefore, it was just needed to look for any significant
correlation between the number of clicks and each of the scores. To
that end, the scores for those users promoting each \emph{URL} were
accumulated and, thus, for each \emph{URL} a number of clicks and
a single total {}``influence'' score were available.

Some caveats should be noted. Firstly, when correlating {}``influence''
with clicks from the \emph{URL} dataset which ignored week limitations,
the velocities employed were those reached by users on August 16,
2009 no matter the date when the \emph{URL} had been published. Certainly,
this is rather unrealistic but consistent with the way in which the
rest of scores were obtained: after all, \emph{PageRank} and \emph{TunkRank}
were computed from a graph crawled well after August 16, and the \emph{retweets}
required to applied the \emph{Influence-Passivity} method were found
across the whole dataset instead of using just the \emph{tweets} predating
the \emph{URLs}.

Secondly, a single empirically found damping factor ($0<\zeta\ll1$)
was applied to compute \emph{velocity}. Proof of concept experiments
showed that dynamic damping (i.e. a constant computed for each week
or day based on the \emph{tweeting} behavior of users during that
period) did not provide better correlation. The same experiments revealed
that a frictionless scenario ($\lyxmathsym{\textgreek{z}}=0$) also
shows a positive correlation between influence and clicks; however,
the correlation was much weaker than when using a positive damping
factor and, thus, such a frictionless model was disregarded.

\subsection*{Experimental findings}

Pearson's \emph{r} was employed to compare \emph{URL} clicks with
accumulated {}``influence''. Certainly, assuming a linear regression
model between a given {}``influence'' score and \emph{URL} visits
can be an oversimplification but, hopefully, it could shed some light
on the relation between such scores and observable events and, besides,
it would make the results of this study comparable to those obtained
by Romero \emph{et al.} who reported $R^{2}$ values.

Table 1 shows the results obtained when comparing the aforementioned
{}``influence'' scores with the visits received by \emph{URLs} in
the dataset ignoring weekly limitations. Coefficients are not too
high but, still, they are significant because of the sample size (22,920
\emph{URLs}). From those results, it seemed that all of the {}``influence''
scores exhibit a positive correlation with \emph{URL} visits; however,
some intriguing questions arise.

First of all, the results greatly departed from those reported by
Romero \emph{et al. }In fact, the correlation found in this study
is much lower than the one reported by those researchers. In addition
to that, the predictive performance of scores such as number of \emph{followers},
\emph{PageRank}, and \emph{I-P} seems to be different than the one
found by them. According to Romero \emph{et al.} \emph{followers<PageRank<I-P}
while Table 1 shows that \emph{I-P<PageRank<followers}.

Needless to say, such differences could be attributed to many factors:
from the datasets themselves to the way in which scores were computed.
Romero \emph{et al.} crawled just \emph{tweets} containing \emph{URLs}
while the dataset employed in this study contained any kind of \emph{tweet}.
While they computed \emph{PageRank} from the \emph{retweeting} graph,
for this study it was computed in a {}``traditional'' way: i.e.
from the \emph{followers} graph. Besides, the way in which \emph{URL}
outliers were considered in both studies could also have distorted
the results. Finally, while they compared average scores of the users
promoting a \emph{URL} with its clicks, accumulated scores were used
for these experiments%
\footnote{During the aforementioned proof of concept experiments it was found
that average scores were worse predictors than accumulated scores.%
}.

All of this would just mean that deeper analyses are needed; nevertheless,
the attentive reader might have noted that a positive correlation
between these {}``influence'' scores and \emph{URL} visits is not
that surprising but, instead, expected. Indeed, the correlation between
the number of \emph{followers} and the clicks received provides a
clue. 

Certainly, algorithms such as \emph{PageRank}, \emph{TunkRank}, or
\emph{Influence-Passivity} are devised in such a way that users with
few \emph{followers} can still achieve rather high scores provided
those few \emph{followers} are {}``influential''. However, this
is not the norm but the exception: most of the users with a high score
also have a large number of \emph{followers}. Hence, if users with
a high \emph{PageRank}, \emph{TunkRank}, \emph{I-P}, or \emph{velocity}
score have lots of \emph{followers}, it is not that strange that the
\emph{URLs} they promote receive more visits than those promoted by
users with lower {}``influence'' scores. After all, they have much
larger audiences and, thus, more visits are to be expected.

Table 2 reveals that a highly significant positive correlation exist
between the number of \emph{followers} and the different {}``influence''
scores. In other words, the accumulated number of \emph{followers}
for the \emph{URLs} must be considered a confounding variable and,
thus, the data must be corrected for it%
\footnote{Although not directly related to the topic of this paper we cannot
fail to urge the reader to consult the recent paper by West \emph{et
al.}\textbf{ }\cite{key-19}.%
}. 

To that end, both clicks and the different {}``influence'' scores
must be divided by the accumulated number of \emph{followers}, in
other words, the expected \emph{audience} for each \emph{URL}. This
way, it would be checked if there exists any correlation between the
probability for a member of a given audience to visit a \emph{URL}
and the portion of the \emph{URL} promoters' influence that member
is responsible for. 

Table 3 shows the results obtained after correcting the data for \emph{audience}.
The results for Influence-Passivity are inconclusive because there
are no significant correlation. The rest of {}``influence'' scores
--namely \emph{PageRank}, \emph{TunkRank}, and \emph{velocity}-- show
significant positive correlations. \emph{Velocity} seems to be the
best predictor, followed by \emph{TunkRank} and, then, \emph{PageRank}.

\begin{table}[h]
\begin{centering}
\begin{tabular}{|c|c|c|c|}
\hline 
\textbf{\small {}``Influence'' score} & \textbf{\small Pearson's }\textbf{\emph{\small r}} & \textbf{\emph{\small $R{}^{\text{2}}$}} & \textbf{\small Significance}\tabularnewline
\hline
\hline 
\textbf{\small Number of }\textbf{\emph{\small followers}} & \textbf{\small 0.26637} & \textbf{\small 0.07095} & \textbf{\small $p\ll0.001$}\tabularnewline
\hline 
\emph{\small Influence (I-P)} & {\small 0.03627} & {\small 0.00132} & {\small $p\ll0.001$}\tabularnewline
\hline 
\emph{\small PageRank} & {\small 0.22381} & {\small 0.05009} & {\small $p\ll0.001$}\tabularnewline
\hline 
\emph{\small TunkRank} & {\small 0.17416} & {\small 0.03033} & {\small $p\ll0.001$}\tabularnewline
\hline 
\emph{\small velocity} & {\small 0.21981} & {\small 0.04832} & {\small $p\ll0.001$}\tabularnewline
\hline
\end{tabular}
\par\end{centering}

\caption{{\small Correlation between different {}``influence'' scores and
clicks received by }\emph{\small URLs}{\small{} in the dataset ignoring
weekly limitations (data was not corrected for audience). Correlation
coefficients are not very high but, given the size of the sample --22,920
}\emph{\small URLs}{\small , all of them are significant ($p\ll0.001$).}}

\end{table}

\begin{table}[h]
\begin{centering}
\begin{tabular}{|c|c|c|c|}
\hline 
\textbf{\small {}``Influence'' score} & \textbf{\small Pearson's }\textbf{\emph{\small r}} & \textbf{\small $R^{2}$} & \textbf{\small Significance}\tabularnewline
\hline
\hline 
\emph{\small Influence (I-P)} & {\small 0.27994} & {\small 0.07836} & {\small $p\ll0.001$}\tabularnewline
\hline 
\emph{\small PageRank} & {\small 0.87284} & {\small 0.76185} & {\small $p\ll0.001$}\tabularnewline
\hline 
\emph{\small TunkRank} & {\small 0.75930} & {\small 0.57653} & {\small $p\ll0.001$}\tabularnewline
\hline 
\emph{\small velocity} & {\small 0.55151} & {\small 0.30417} & {\small $p\ll0.001$}\tabularnewline
\hline
\end{tabular}
\par\end{centering}

\caption{{\small Correlation between the accumulated number of }\emph{\small followers}{\small{}
and the rest of {}``influence'' scores using the information in
the dataset ignoring weekly limitations. As it can be seen there exists
a significant ($p\ll0.001$) positive correlation between the number
of }\emph{\small followers}{\small{} and the {}``influence'' scores.
}\emph{\small Influence-Passivity}{\small{} seems to be the method
less sensitive to the number of }\emph{\small followers}{\small{} and
}\emph{\small PageRank}{\small{} the most sensitive.}}

\end{table}

\begin{table}[h]
\begin{centering}
\begin{tabular}{|c|c|c|c|}
\hline 
\textbf{\small {}``Influence'' score} & \textbf{\small Pearson's }\textbf{\emph{\small r}} & \textbf{\small $R^{2}$} & \textbf{\small Significance}\tabularnewline
\hline
\hline 
\emph{\small Influence (I-P)} & {\small -0.01021} & {\small 0.00010} & {\small Non-significant}\tabularnewline
\hline 
\emph{\small PageRank} & {\small 0.04399} & {\small 0.00194} & {\small $p\ll0.001$}\tabularnewline
\hline 
\emph{\small TunkRank} & {\small 0.13550} & {\small 0.01836} & {\small $p\ll0.001$}\tabularnewline
\hline 
\textbf{\emph{\small velocity}} & \textbf{\small 0.26532} & \textbf{\small 0.07039} & \textbf{\small $p\ll0.001$}\tabularnewline
\hline
\end{tabular}
\par\end{centering}

\caption{{\small Correlation between different {}``influence'' scores and
clicks received by }\emph{\small URLs}{\small{} in the dataset ignoring
weekly limitations after correcting for the confounding variable }\emph{\small audience}{\small{}
(i.e. scores and clicks were divided by the accumulated number of
}\emph{\small followers }{\small of the users promoting the }\emph{\small URLs}{\small ).
All of the scores, except for }\emph{\small I-P}{\small , show significant
positive correlations.}}

\end{table}

It should be remembered that all of these results were obtained from
the first \emph{URL} dataset which did not take into consideration
weekly limitations and, because of that, \emph{velocity} scores were
those reached by users on August 16, 2009. Another set of results
was obtained by using the second dataset, comprising \emph{URLs} which
appeared in one single week. 

For those experiments, three different \emph{velocity} scores were
employed: \emph{(1)} \emph{velocities} reached on August 16, 2009;
\emph{(2)} \emph{velocities} computed at the end of each week; and
\emph{(3)} \emph{velocities} computed at the end of the prior week.
It is easy to see that the third {}``flavor'' is the closest one
to a real-time application.

Needless to say, the correlation coefficients reported in Table 4
were obtained by averaging the coefficients found for each week (cf.
Cramer \& Howitt \cite{key-21}, p.40) while the significance was
computed according to the average sample size (349 \emph{URLs} per
week). These results are pretty consistent with those of Table 3:
the correlation between \emph{Influence-Passivity} and clicks is again
non-significant; the rest of scores exhibit a significant positive
correlation with \emph{URL} visits; and, again, \emph{velocity} is
the best predictor. 

On a side note, \emph{velocities} computed on the week when \emph{URLs}
were published are slightly better predictors than \emph{velocities}
computed the week before. This would be of course expected if velocity
in \emph{Twitter} was a valid proxy measure for influence.

\begin{table}[h]
\begin{centering}
\begin{tabular}{|c|c|c|c|}
\hline 
\textbf{\small {}``Influence'' score} & \textbf{\small Pearson's }\textbf{\emph{\small r}} & \textbf{\small $R^{2}$} & \textbf{\small Significance}\tabularnewline
\hline
\hline 
\emph{\small Influence (I-P)} & {\small 0.06806} & {\small 0.00463} & {\small Non-significant}\tabularnewline
\hline 
\emph{\small PageRank} & {\small 0.25418} & {\small 0.06461} & {\small $p\ll0.001$}\tabularnewline
\hline 
\emph{\small TunkRank} & {\small 0.29921} & {\small 0.08952} & {\small $p\ll0.001$}\tabularnewline
\hline 
\emph{\small velocity}{\small{} (August 16)} & {\small 0.35464} & {\small 0.12577} & {\small $p\ll0.001$}\tabularnewline
\textbf{\emph{\small velocity}}\textbf{\small{} (on week)} & \textbf{\small 0.37735} & \textbf{\small 0.14240} & \textbf{\small $p\ll0.001$}\tabularnewline
{\small velocity (prior week)} & {\small 0.37437} & {\small 0.14015} & {\small $p\ll0.001$}\tabularnewline
\hline
\end{tabular}
\par\end{centering}

\caption{{\small Average correlation coefficients between different {}``influence''
scores and clicks received by }\emph{\small URLs}{\small{} in the dataset
with weekly limitations. Both clicks and scores were corrected for
the confounding variable }\emph{\small audience}{\small . Reported
coefficients were obtained by averaging the coefficients computed
for each week.}}

\end{table}

\section*{Case study --Real-time detection of trending users by using acceleration}

Perhaps one of the most direct applications of the new method described
in this paper is to detect \emph{trending} users; that is, those users
reaching high \emph{velocities} and who can be of interest for an
audience that is still unaware of them. 

The most straightforward way of finding such users would be computing
the difference between the users' current \emph{velocities} and their
previous ones to, then, order them by decreasing \emph{acceleration}. 

Nevertheless, by doing this there exists the risk of obtaining many
users with high \emph{accelerations} in absolute terms but rather
low, even irrelevant, in relative terms (that would be the case of
the most popular users, for instance). 

To avoid this problem those users with a relative increase in \emph{velocity}
below a certain threshold (e.g. 10\%) could be filtered out, and then
the remaining users would be ordered by decreasing \emph{acceleration}.
This method was applied to the \emph{Twitter} dataset to obtain a
list of trending users for each week from January 26 to August 16,
2009. 

A thorough evaluation of the quality of those results was out of the
scope of this study; still, an informal analysis of the top ranked
\emph{trending} users was conducted. To that end, the \emph{tweets}
mentioning the top-5 \emph{trending} users for each week were obtained,
and the most common phrases within them were obtained. Those phrases
and the name of the user --generally a celebrity-- were used to query
a search engine. From the obtained results it was possible in virtually
all of the cases to determine one or more actual events involving
the user, and explaining the sudden increase in \emph{velocity}. 

Tables 5 to 9 show a summary of that informal evaluation; as it can
be seen, the results obtained by applying the technique proposed in
this paper seem highly promising.

{\scriptsize }%
\begin{table}[h]
{\scriptsize }\begin{tabular}{|l|l|>{\raggedright}p{0.2\columnwidth}|>{\raggedright}p{0.5\columnwidth}|}
\hline 
\textbf{\scriptsize Week} & \textbf{\scriptsize Twitter user} & \textbf{\scriptsize Real name} & \textbf{\scriptsize Explanation and most frequent phrases}\tabularnewline
\hline
\hline 
{\scriptsize Feb. 1, 2009 } & {\scriptsize stephenfry } & {\scriptsize Stephen Fry (English actor, writer, comedian, TV presenter
and film director) } & {\scriptsize Stephen Fry was to appear on February 2, 2009 at an Apple
Store in London to present his new audiobook.}{\scriptsize \par}

\textbf{\scriptsize apple store}\tabularnewline
\hline 
{\scriptsize Feb. 8, 2009 } & {\scriptsize wossy } & {\scriptsize Jonathan Ross (English TV and radio presenter) } & {\scriptsize (1) Ross, host for the 2009 edition of the Bafta Awards
held on February 8, 2009, asked his followers for a word to insert
during the ceremony; the chosen word was {}``salad''. (2) On February
6, 2009 Tom Jones and Anna Friel, among others, visited Friday Night
with Jonathan Ross.}{\scriptsize \par}

\textbf{\scriptsize word salad, use word, good luck baftas, bafta
word, twitter word, tom jones, anna friel}\tabularnewline
\hline 
{\scriptsize Feb. 15, 2009} & {\scriptsize lancearmstrong } & {\scriptsize Lance Armstrong (American professional road racing cyclist)} & {\scriptsize Lance Armstrong's time-trial bike was stolen on February
14, 2009 before the first stage of the Tour of California.}{\scriptsize \par}

\textbf{\scriptsize stolen tt bike, time trial bike, bike stolen,
tour california}\tabularnewline
\hline 
{\scriptsize Feb. 22, 2009} & {\scriptsize the\_real\_shaq } & {\scriptsize Shaquille O'Neal (American professional basketball player) } & {\scriptsize (1) Mentions to the All-Star Game of the last weekend.
(2) On February 20, 2009 O'Neal suggested to all of his followers
to introduce themselves because they are connected in the, Shaquille
wording, {}``Twitteronia''.}{\scriptsize \par}

\textbf{\scriptsize star game, twitteronia, public come say hi, twitteronia
connect, congrats mvp}\tabularnewline
{\scriptsize Mar. 1, 2009} &  &  & {\scriptsize On February 24, 2009 Shaquille suggested his followers
to meet him in a mall to get two tickets. }{\scriptsize \par}

\textbf{\scriptsize fashion sq mall, touches gets 2 tickets}\tabularnewline
\hline 
{\scriptsize Mar. 8, 2009} & {\scriptsize iamdiddy } & {\scriptsize Sean Combs (American record producer, rapper, actor,
and fashion designer) } & \emph{\scriptsize Unknown.}{\scriptsize \par}

\textbf{\scriptsize bad boy, positive energy, first time, god bless}\tabularnewline
\hline
\end{tabular}{\scriptsize \par}

{\scriptsize \caption{{\small Top }\emph{\small trending}{\small{} users (ranking at the
first position) found during February and early March 2009. The dates
reported are those of the last day in the corresponding week. The
third column identifies the individual or company and provides a short
description. The last column provides an explanation for that user
being }\emph{\small trending}{\small{} on that week plus a number of
the most frequent phrases found in the }\emph{\small tweets}{\small{}
mentioning the user during that week. As it can be seen, most of the
times the }\emph{\small tweets}{\small{} dealt with the actual events
in which the user was involved.}}
}
\end{table}
{\scriptsize \par}

{\scriptsize }%
\begin{table}[h]
{\scriptsize }\begin{tabular}{|l|l|>{\raggedright}p{0.2\columnwidth}|>{\raggedright}p{0.5\columnwidth}|}
\hline 
\textbf{\scriptsize Week} & \textbf{\scriptsize Twitter user} & \textbf{\scriptsize Real name} & \textbf{\scriptsize Explanation and most frequent phrases}\tabularnewline
\hline
\hline 
{\scriptsize Mar. 15, 2009} & {\scriptsize theellenshow } & {\scriptsize Ellen DeGeneres (American stand-up comedian, TV host
and actress) } & {\scriptsize Ellen DeGeneres joined Twitter on March 10, on March
11 she was to appear at the Jay Leno show and she made a public appeal
to get followers. In fact, most of the tweets mentioning @theellenshow
were retweets of her original one: }\emph{\scriptsize \textquotedbl{}tweet
\& call everyone you know \& tell them to follow me- I want to see
how many I can get by the time I'm on Leno tonight.\textquotedbl{} }{\scriptsize \par}

\textbf{\scriptsize leno tonight, tell follow, see many, many can
get, call everyone}\tabularnewline
\hline 
{\scriptsize Mar. 22, 2009} & {\scriptsize lancearmstrong } & {\scriptsize Lance Armstrong (American professional road racing cyclist) } & {\scriptsize (1) On March 23, 2009 Lance Armstrong broke his collarbone
in a crash during a race in Spain and had to face surgery. (2) On
March 17, 2009 Lance Armstrong was required by French anti-doping
agency to provide a hair sample. }{\scriptsize \par}

\textbf{\scriptsize good luck, get well soon, best wishes, anti-doping,
hope ok, recovery just, luck surgery, broken clavicle}\tabularnewline
\hline 
{\scriptsize Mar. 29, 2009} & {\scriptsize macheist} & {\scriptsize MacHeist (website reselling Mac OS X shareware)} & {\scriptsize On March 24, 2009 the MacHeist 3 bundle was revealed
in a live show. }{\scriptsize \par}

\textbf{\scriptsize bundle reveal show, 3 bundle, buy bundle }\tabularnewline
{\scriptsize Apr. 5, 2009} &  &  & {\scriptsize On March 25, 2009 the MacHeist 3 Bundle was on sale featuring
12 popular Mac applications normally valued at over \$900 for just
\$39. }{\scriptsize \par}

\textbf{\scriptsize macheist 3 bundle, mac apps, just \$39, mac apps
worth \$900+, 12 top mac apps}\tabularnewline
\hline 
{\scriptsize Apr. 12, 2009} & {\scriptsize joeymcintyre} & {\scriptsize Joey McIntyre (American singer-songwriter and actor,
part of the band New Kids on the Block)} & {\scriptsize The @joeymcintyre account was created on April 9, 2009
so, probably, that's the reason for it's sudden popularity. Most of
the topics seem to be related to the \textquotedbl{}Full Service\textquotedbl{}
summer tour in which NKOTB were involved. }{\scriptsize \par}

\textbf{\scriptsize summer tour, happy easter, full service, easter
bunny}\tabularnewline
\hline 
{\scriptsize Apr. 19, 2009} & {\scriptsize jordanknight} & {\scriptsize Jordan Knight (American singer-songwriter, part of the
band New Kids on the Block)} & \emph{\scriptsize (Tentative)}{\scriptsize{} Jordan Knight joined Twitter
on April 14, 2009 and started to be addressed by fans with the rest
of members of NKOTB. }{\scriptsize \par}

\textbf{\scriptsize dannywood, jonathanrknight, donniewahlberg dannywood,
joeymcintyre donniewahlberg}\tabularnewline
\hline 
{\scriptsize May 3, 2009} & {\scriptsize jonasbrothers} & {\scriptsize Jonas Brothers (American pop boy band)} & {\scriptsize On April 30, 2009 it was announced that Jonas Brothers
would be participating in a series of live web chats starting on May
7. }{\scriptsize \par}

\textbf{\scriptsize may 7th, live web chat may, question jonaslive}\tabularnewline
\hline
\end{tabular}{\scriptsize \par}

{\scriptsize \caption{{\small Top }\emph{\small trending}{\small{} users for March, April
and early May. No data is provided for the week ending on April 26,
2009 because the dataset lacks several days on that week and, hence,
all of the users lost }\emph{\small velocity}{\small .}}
}
\end{table}
{\scriptsize \par}

{\scriptsize }%
\begin{table}[h]
{\scriptsize }\begin{tabular}{|l|l|>{\raggedright}p{0.2\columnwidth}|>{\raggedright}p{0.5\columnwidth}|}
\hline 
\textbf{\scriptsize Week} & \textbf{\scriptsize Twitter user} & \textbf{\scriptsize Real name} & \textbf{\scriptsize Explanation and most frequent phrases}\tabularnewline
\hline
\hline 
{\scriptsize May 10, 2009} & {\scriptsize jordanknight} & {\scriptsize Jordan Knight (American singer-songwriter, part of the
band New Kids on the Block)} & \emph{\scriptsize (Tentative)}{\scriptsize{} Jordan tweeted }\emph{\scriptsize {}``Tink!
is the imaginary sound of my eyelids springing open when I wake up''.}{\scriptsize{} }{\scriptsize \par}

\textbf{\scriptsize today show, joeymcintyre dannywood, donniewahlberg
jonathanrknight, jonathanrknight joeymcintyre, tink sound}\tabularnewline
\hline 
{\scriptsize May 17, 2009} & {\scriptsize onlinesystem} & {\scriptsize Online System (¿online marketer?)} & \emph{\scriptsize (Tentative) }{\scriptsize The user seems to be an
aggressive marketer promoting systems to earn money throw affiliate
marketing, virtually all of the tweets seem to be users reporting
the increase in followers they got using the user's method. }{\scriptsize \par}

\textbf{\scriptsize followers using twitter, new followers added,
20 new followers added}\tabularnewline
\hline 
{\scriptsize May 24, 2009} & {\scriptsize jonasbrothers} & {\scriptsize Jonas Brothers (American pop boy band)} & {\scriptsize (1) {}``Paranoid'' was the first single from their
then new album; the video premiered on May 23, 2009. (2) Jonas Brothers
play a little role in the movie {}``Night at the Museum: Battle of
the Smithsonian,'' sequel to the film {}``Night at the Museum,''
which was released in theaters on May 22, 2009. }{\scriptsize \par}

\textbf{\scriptsize music video, night museum 2, music video paranoid }\tabularnewline
{\scriptsize May 31, 2009} &  &  & {\scriptsize On May 28, 2009 another live web chat with the Jonas
Brothers was held. }{\scriptsize \par}

\textbf{\scriptsize web chat, web chat may 28th, new album, new songs,
night museum 2 }\tabularnewline
\hline 
{\scriptsize Jun. 7, 2009} & {\scriptsize mileycyrus} & {\scriptsize Miley Cyrus (American actress and pop singer)} & {\scriptsize {}``The Climb,'' performed by Miley Cyrus for {}``Hannah
Montana: The Movie,'' won at the 2009 MTV Movie Awards held on May
31, 2009 in the category {}``Best Song from a Movie''. }{\scriptsize \par}

\textbf{\scriptsize hannah montana, mtv movie awards, best song, congrats
award, song climb, best song movie, congratulations }\tabularnewline
\hline 
{\scriptsize Jun. 14, 2009} & {\scriptsize peterfacinelli} & {\scriptsize Peter Facinelli (American actor)} & {\scriptsize Peter Facinelli made a bet with Rob DeFranco that he
could get 500,000 followers on Twitter by June 19. If Facinelli wasn't
able to win the bet he should have to give DeFranco his Twilight chair.
However, if he won the bet DeFranco should have to walk down Hollywood
Blvd. in a bikini singing {}``All the Single Ladies''. }{\scriptsize \par}

\textbf{\scriptsize win bet, single ladies, 500,000 followers, rob
defranco, next week, bikini dance }\tabularnewline
\hline
\end{tabular}{\scriptsize \par}

{\scriptsize \caption{{\small Top }\emph{\small trending}{\small{} users for May and mid
June 2009.}}
}
\end{table}
{\scriptsize \par}

{\scriptsize }%
\begin{table}[h]
{\scriptsize }\begin{tabular}{|l|l|>{\raggedright}p{0.2\columnwidth}|>{\raggedright}p{0.5\columnwidth}|}
\hline 
\textbf{\scriptsize Week} & \textbf{\scriptsize Twitter user} & \textbf{\scriptsize Real name} & \textbf{\scriptsize Explanation and most frequent phrases}\tabularnewline
\hline
\hline 
{\scriptsize Jun. 21, 2009} & {\scriptsize perezhilton} & {\scriptsize Perez Hilton (American blogger and TV personality)} & {\scriptsize On June 17, 2009 Hilton used Twitter to claim assault
by the Black Eyed Peas member will.i.am and his security guards. }{\scriptsize \par}

\textbf{\scriptsize call police, black eyed peas, assaulted will,
security wards}\tabularnewline
\hline 
{\scriptsize Jun. 28, 2009} & {\scriptsize songzyuuup} & {\scriptsize Trey Songz fan page (Trey Songz is an American recording
artist, producer and actor)} & {\scriptsize Trey Songz attended and performed at the BET Awards ceremony
held on June 28, 2009. }{\scriptsize \par}

\textbf{\scriptsize bet awards, love trey, good bet awards, loved
performance}\tabularnewline
\hline 
{\scriptsize Jul. 5, 2009} & {\scriptsize mileycyrus} & {\scriptsize Miley Cyrus (American actress and pop singer)} & {\scriptsize (1) Miley Cyrus starred in {}``Hanna Montana: The Movie''
which as of July 2009 was still on theaters. (2) Cyrus started shooting
the movie {}``The Last Song'' on June 15, 2009. }{\scriptsize \par}

\textbf{\scriptsize hannah montana, hanna montana movie, last song}\tabularnewline
\hline 
{\scriptsize Jul. 12, 2009} & {\scriptsize songzyuuup} & {\scriptsize Trey Songz fan page (Trey Songz is an American recording
artist, producer and actor)} & {\scriptsize On June 2009 Trey Songz released a mixtape titled {}``Anticipation''
through his blog before releasing this third album. }{\scriptsize \par}

\textbf{\scriptsize trey songz, anticipation album, listening anticipation,
mixtape anticipation}\tabularnewline
\hline 
{\scriptsize Jul. 19, 2009} & {\scriptsize jordanknight} & {\scriptsize Jordan Knight (American singer-songwriter, part of the
band New Kids on the Block)} & {\scriptsize A concert by New Kids on the Block was live webcasted
on July 17, 2009. }{\scriptsize \par}

\textbf{\scriptsize webcast, jordan girl, love u, thank u, full service,
good luck, luv u, love ya, miss u}\tabularnewline
\hline 
{\scriptsize Jul. 26, 2009} & {\scriptsize myfabolouslife} & {\scriptsize Fabolous (American recording artist)} & \emph{\scriptsize (Tentative) }{\scriptsize Fan comments about the
official remix of {}``Throw It in the Bag'' featuring rapper Drake.
The remix was released on August 18, 2009. }{\scriptsize \par}

\textbf{\scriptsize throw bag, throw bag remix, ft drake, remix official}\tabularnewline
\hline
\end{tabular}{\scriptsize \par}

{\scriptsize \caption{{\small Top }\emph{\small trending}{\small{} users for June and July,
2009.}}
}
\end{table}
{\scriptsize \par}

{\scriptsize }%
\begin{table}[h]
{\scriptsize }\begin{tabular}{|l|l|>{\raggedright}p{0.2\columnwidth}|>{\raggedright}p{0.5\columnwidth}|}
\hline 
\textbf{\scriptsize Week} & \textbf{\scriptsize Twitter user} & \textbf{\scriptsize Real name} & \textbf{\scriptsize Explanation and most frequent phrases}\tabularnewline
\hline
\hline 
{\scriptsize Aug. 2, 2009} & {\scriptsize paulaabdul} & {\scriptsize Paula Abdul (American pop singer, record producer, dancer,
actress and TV personality)} & {\scriptsize During the 2000s Paula Abdul acted as judge on the TV
contest {}``American Idol.'' On July 17, 2009 her manager announced
that she'd leave the show if producers didn't step up a new deal.
It wasn't until August 4, 2009 that Paula definitely that she wouldn't
return to {}``Idol.'' In the mean time many followers tweet their
support for Abdul using the hashtag }\emph{\scriptsize \#keeppaula}{\scriptsize . }{\scriptsize \par}

\textbf{\scriptsize \#keeppaula, will continue}\tabularnewline
\hline 
{\scriptsize Aug. 9, 2009} & {\scriptsize adamlambert} & {\scriptsize Adam Lambert (American singer, songwriter, and actor)} & {\scriptsize (1) }\emph{\scriptsize (Tentative)}{\scriptsize{} NOH8
Campaign was a silent protest photo project against California Proposition
8; it seems that Lambert fans were campaigning to get their idol taking
part of the project. (2) On August 9, 2009 Adam Lambert won a Teen
Choice Award. }{\scriptsize \par}

\textbf{\scriptsize wewant 4noh8, noh8campaign, noh8campaign wewant,
teen choice awards}\tabularnewline
\cline{4-4} 
{\scriptsize Aug. 16, 2009} &  &  & {\scriptsize (1) On August 13, 2009 Lambert answered fan questions
by means of Twitter in a so-called {}``Twitter party.'' (2) On August
9, 2009 creative director of ELLE tweet about having Adam Lambert,
among others, to the creative photo shooting for the next edition
of the magazine. (3) On August 9, 2009 Adam Lambert won a Teen Choice
Award. }{\scriptsize \par}

\textbf{\scriptsize twitter party, elle shoot, details elle shoot,
teen choice awards}\tabularnewline
\hline
\end{tabular}{\scriptsize \par}

{\scriptsize \caption{{\small Top }\emph{\small trending}{\small{} users for August, 2009.}}
}
\end{table}
{\scriptsize \par}

\section*{Conclusions}

This paper has described a new method to compute \emph{Twitter} \emph{influence}
based on a physical metaphor which has got a number of advantages
over commonly applied techniques. 

First, it does not rely on the \emph{Twitter} user graph which is
costly to crawl, just provides static \emph{snapshots} of a rapidly
evolving network, and does not represent actual user interactions.
Instead, the new method just requires the\emph{ }streamline of \emph{tweets}
to detect user \emph{mentions}. 

Second, it can be applied in near real-time and provides a natural
way to detect \emph{trending} or emerging users. Some anecdotal evidence
on the quality of this approach has been provided.

A number of experiments were conducted to check whether the new \emph{velocity
}score actually correlates with \emph{influence}. Results from those
experiments have been reported, revealing that most of the commonly
applied scores such as the number of \emph{followers}, or \emph{PageRank},
and recently proposed ones such as \emph{TunkRank}, or \emph{Influence-Passivity},
certainly exhibit a positive correlation with website visits. 

However, it has also been shown that the number of \emph{followers}
is a confounding variable which must be accounted for. Therefore,
it is not the total number of visits and the different {}``influence''
scores which have to be correlated but, instead, the probability of
a user visiting a promoted \emph{URL} and the proportion of the promoter's
influence a single user is responsible for.

After correcting the data for the audience, it was revealed that all
of the {}``influence'' scores except for one --namely, \emph{PageRank},
\emph{TunkRank}, and \emph{velocity}-- exhibit positive correlation
with user clicks and, thus, with \emph{influence} in the sense of
{}``attention gathering''. \emph{Velocity}, the score inferred by
the method proposed in this paper, was by a large margin the best
predictor of user clicks.

The only score not showing significant correlation was \emph{Influence-Passivity}.
There exist, however, a number of reasons for this inconclusive result.
The main one is, in all probability, the sparseness of the \emph{retweet}
graph obtained from the dataset because of the strict requirements
of the \emph{Influence-Passivity} method (i.e. a user has not only
to \emph{follow} another one but \emph{retweet} some of his messages).

Hence, this study makes a number of contributions. \emph{(1) }It adds
to the general understanding of the concept of \emph{influence} in
\emph{OSNs }and its relation to {}``attention gathering''; \emph{(2)}
it has exposed the caveat due to the confounding nature of \emph{audience}
in this scenario; \emph{(3)} it has shown how centrality measures
can be used as rather good predictors of \emph{influence}; and \emph{(4)}
it has described a new method that outperforms them with regards to
\emph{influence} scoring, and that can be applied in real-time to
rank users and to detect emerging\emph{ }{}``influentials''. In
this sense, an interesting future line of work would be studying the
feasibility of adapting this new model to \emph{tweets} themselves
to detect \emph{trending topics} and compare its performance with
\emph{Twitter}'s own implementation.

\end{document}